# Observing a quantum magnetic effect in $CaYAl_3O_7$-X particles (X=Ni, Fe, Co)

Luana Hildever[1,2], José Laurentino[1,2], José Araújo[2,3], Francisco Estrada[4], and José Holanda[1,2,3*]

[1]Programa de Pós-Graduação em Engenharia Física, Universidade Federal Rural de Pernambuco, 54518-430, Cabo de Santo Agostinho, Pernambuco, Brazil

[2]Group of Optoelectronics and Spintronics, Universidade Federal Rural de Pernambuco, 54518-430, Cabo de Santo Agostinho, Pernambuco, Brazil

[3]Unidade Acadêmica do Cabo de Santo Agostinho, Universidade Federal Rural de Pernambuco, 54518-430, Cabo de Santo Agostinho, Pernambuco, Brazil

[4]Facultad de Biologia, Universidad Michoacana de San Nicolas de Hidalgo, Av. F. J. Mujica s/n Cd. Universitaria, Morelia, Michoacian, México.

## Abstract

Yttrium calcium aluminate, with the formula $CaYAl_3O_7$, has been extensively researched due to its remarkable luminescent properties when doped or co-doped. Additionally, it exhibits exceptional piezoelectric properties at high temperatures. However, the potential magnetic properties this material can acquire through doping or co-doping have largely gone unexplored until now. In this innovative study, we investigate the quantum magnetic characteristics of yttrium calcium aluminate particles doped with transition metals such as cobalt, iron, and nickel. Our magnetic measurements of hysteresis curves in the macroscopic regime reveal intriguing magnetic characteristics that have not been documented in the literature. Given the high applicability of this material in biological microelectromechanical systems, we conducted a detailed study to demonstrate these findings.

Corresponding author: * joseholanda.silvajunior@ufrpe.br

Studying the optical and magnetic properties of particles is essential for advancing our understanding of various technological applications. The ability to control these characteristics is crucial for developing innovative materials and technologies. Research into the optical and magnetic properties of materials has led to a broader understanding of characteristics that are not yet fully understood. In this context, materials doped with rare earth ions and transition metals show significant promise, as doping can significantly alter their properties, bestowing them with desirable optical and magnetic characteristics. One notable material with exceptional properties is the host matrix $ABC_3O_7$, which crystallizes in the tetragonal system with the space group P421m, where A represents Ca, Sr, or Ba; B represents La or Gd; and C represents Al or Ga. Specifically, the single crystal $CaYAl_3O_7$ (CYAM) belongs to the non-centrosymmetric melilite family and has garnered considerable interest due to its numerous advantages, including good physicochemical stability, the availability of raw materials, and straightforward synthesis conditions. Additionally, CYAM exhibits remarkable luminescence properties when doped or co-doped with rare earth ions and demonstrates piezoelectric properties under high-temperature conditions. These characteristics are crucial for various photonic applications, such as temperature sensors, the construction of white light LEDs, and electronic devices, including actuators, ultrasonic devices, and electrical transformers.

The luminescent properties of CYAM are typically studied using undoped powder samples. Morphological analyses often reveal particles with regular shapes. Photoluminescence (PL) emission and excitation spectra, obtained using synchrotron radiation in the ultraviolet and vacuum ultraviolet spectral regions, indicate that the bandgap energy of CYAM is approximately 6.8 eV or lower. A low-intensity emission band around 4.4 eV, observed when excited at 6.5 eV, is attributed to self-trapped excitons generated during vacuum ultraviolet excitation. Three main emission bands have been identified, peaking at 2.57, 2.94, and 3.23 eV, which account for the majority of PL and radioluminescence (RL) emission in CYAM. To date, calcium yttrium aluminum oxide phosphors doped with rare earth ions have been extensively studied to achieve full-color red and blue emissions. Additionally, research on exciton luminescence in single crystals (SCs) and single crystal films (SCFs) of $YAlO_3$ has demonstrated that the radiative annihilation of excitons in SCFs primarily occurs at regular sites within the perovskite lattice, resulting in luminescence with a peak at $\lambda_{max} = 295$ nm.

In yttrium (Y)-based aluminate systems, emission bands have been identified within an energy range of 2.53 to 3.49 eV [21, 22]. Exciton emissions typically involve two or more bands, and their positioning is influenced by several factors, including the band gap of the oxide, the distribution of cations across different lattice sites, the chosen synthesis method, and the measurement temperature [22-24]. While CYAM is a frequently studied structure due to its unique characteristics, there have been no reports on the magnetic properties of these particles when doped with transition metals. In this study, we observed magnetic features for the first time in yttrium calcium aluminate particles with the composition $CaYAl_3O_7$, which were doped with cobalt (Co), iron (Fe), and nickel (Ni). A diagram of the unit cell structure of $CaYAl_3O_7$, showing the replacement of Ca/Y ions with Co, Fe, or Ni ions, is presented in **Fig. 1**.

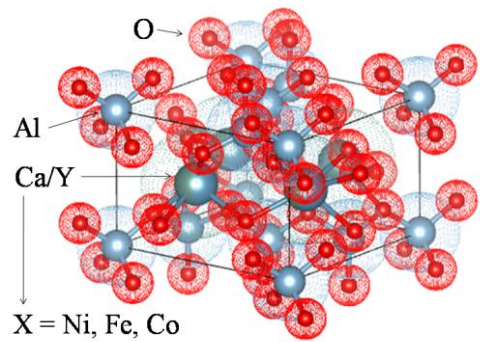

**Figure 1.** The unit cell structure of $CaYAl_3O_7$, where Ca and Y ions are replaced by Co, Fe, or Ni ions.

The ability to manipulate particle size, shape, composition, and properties allows for the creation of materials with unique characteristics tailored to specific applications. There are several methods for particle synthesis, each with its own advantages and limitations depending on the desired properties. Among these methods are the Pechini method, combustion synthesis, and the sol-gel method. The Pechini method is particularly recognized for its capability to produce particles with specific sizes and compositions while accommodating the incorporation of various elements into the particle structure. For our samples, we opted to use this method, applying a 10% transition metal doping fraction. In **Fig. 2 (a)**, we present typical X-ray diffraction measurements of our samples. We conducted a detailed analysis of the structure of our samples to compare the pure $CaYAl_3O_7$ compositions with those doped with Ni, Fe, and Co against the ICSD pattern #09438. This analysis revealed that the Ca/Y ions were replaced by the transition metal ions (Ni, Fe, and Co), as shown in **Fig. 2 (a)**. Additionally, **Figs. 2 (b)** and **(c)** provide an overview of the main techniques employed in our experiments: Confocal Raman Microscopy (**Fig. 2 (b)**) and Vibrating Sample Magnetometry (**Fig. 2 (c)**).

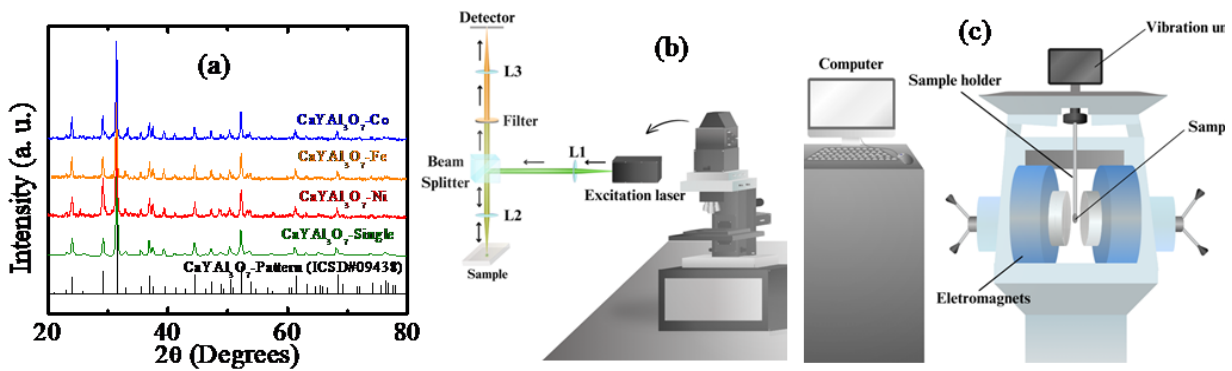

**Figure 2. (a)** X-ray diffraction measurements of pure and Ni-, Fe-, and Co-doped $CaYAl_3O_7$ samples, compared with the ICSD#09438 pattern. Sketches of the Confocal Raman Microscopy **(b)** and Vibrant Sample Magnetometry **(c)** techniques.

The structural properties of a material can influence its optical properties. To investigate this, we conducted Confocal Raman Microscopy measurements using the setup shown in **Fig. 2 (b)**. This technique combines the high spatial resolution of confocal microscopy with the analytical capabilities of Raman spectroscopy. During the process, a laser beam is directed at the sample, interacting with its molecules. The scattered light reveals changes in the vibrational energy of these molecules, providing insights into their chemical composition. This method involves the use of a high-intensity laser, which induces inelastic scattering of light known as Raman scattering. Each unique scattering event generates a characteristic molecular Raman spectrum, allowing us to identify the different chemical components present in the particles. In our measurements, we utilized two laser lines with wavelengths of 532 nm and 633 nm. The particles were deposited on a microscope slide for analysis. In **Fig. 3**, we present the measurements performed on our samples, which include three types of particles: $CaYAl_3O_7$-Ni, $CaYAl_3O_7$-Fe, and $CaYAl_3O_7$-Co, considering both laser wavelengths. Each Raman spectrum was obtained from the respective points marked with an 'X' in the microscopy maps shown in **Fig. 3**. For all $CaYAl_3O_7$ samples doped with transition metals, the observed peaks correspond to the energies of $CaYAl_3O_7$, along with the resonant energies of the transition metals. The Raman signals observed for $CaYAl_3O_7$ are consistent with those described in the literature.

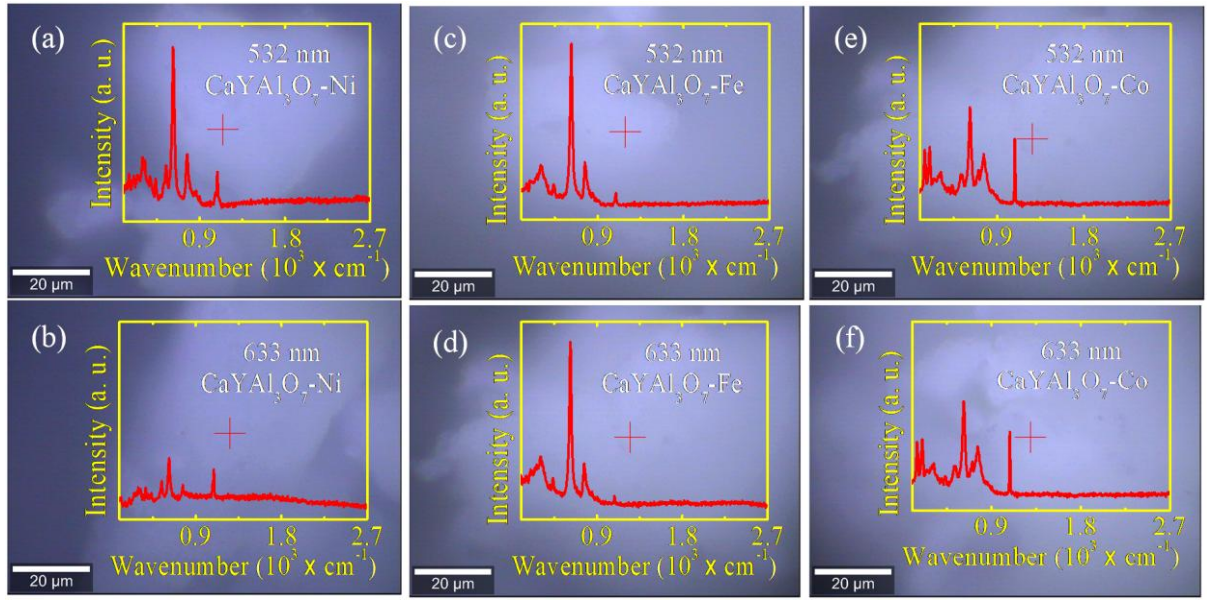

**Figure 3.** Confocal Raman Microscopy measurements of the $CaYAl_3O_7$-X samples (X = Ni, Fe, and Co). Specifically for the two laser lines, the following measurements were performed: $CaYAl_3O_7$-Ni, **(a)** 532 nm and **(b)** 633 nm; $CaYAl_3O_7$-Fe, **(c)** 532 nm and **(d)** 633 nm; $CaYAl_3O_7$-Co, **(e)** 532 nm and **(f)** 633 nm.

The $CaYAl_3O_7$ material exhibits excellent optical characteristics, as evidenced by the spectra presented in **Fig. 3**. Notably, when doped with transition metals that replace the Ca/Y ions, it maintains its optical properties [7-17]. In this context, we investigated whether the doped materials exhibited any magnetic properties. To our surprise, we found clear evidence of magnetic effects through hysteresis curve measurements, as shown in **Fig. 4**. Our observations indicate that the doped materials demonstrated increased coercivity and remanence, following the order: $CaYAl_3O_7$-Co, $CaYAl_3O_7$-Fe, and $CaYAl_3O_7$-Ni, as illustrated in **Figs. 4 (a)**, **(b)**, and **(c)**. The successful positioning of transition metal ions within the crystal structures of the doped materials strengthened the spin interactions of the magnetic ions, making these interactions observable through macroscopic measurements such as hysteresis curves.

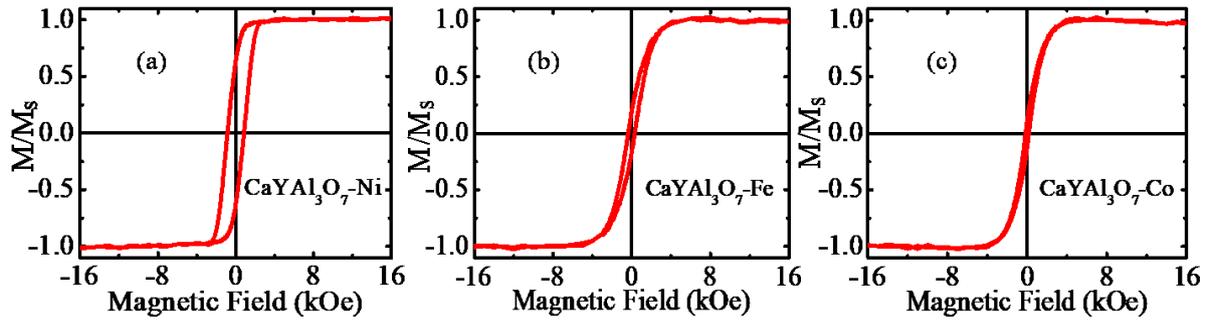

**Figure 4**. Hysteresis curve measurements made using the Vibrating Sample Magnetometry technique for the three materials: **(a)** $CaYAl_3O_7$-Ni, **(b)** $CaYAl_3O_7$-Fe, and **(c)** $CaYAl_3O_7$-Co.

We present new findings on the magnetic properties of $CaYAl_3O_7$ when doped with transition metal ions such as Co, Fe, or Ni. The optical properties of $CaYAl_3O_7$ remained unchanged [1-3, 7-17], as did the properties of the transition metals themselves [4-6, 31, 32]. However, upon doping with these transition metals, $CaYAl_3O_7$ acquired notable magnetic properties, which we demonstrated through magnetic measurements of hysteresis curves. This observed effect arises from the interactions between the spins of the transition metal ions, which are purely quantum in nature. This is significant because we were able to detect this effect through macroscopic magnetic measurements. The doped materials studied here show great promise for applications in biological microelectromechanical systems.

**Acknowledgements**

This research was supported by Conselho Nacional de Desenvolvimento Científico e Tecnológico (CNPq) with Grant Number: 309982/2021-9, Coordenação de Aperfeiçoamento de Pessoal de Nível Superior (CAPES) with Grant Number: PROAP2024UFRPE, and Fundação de Amparo à Ciência e Tecnologia do Estado de Pernambuco (FACEPE) with Grant Number: APQ-1397-3.04/24. The Confocal Raman Magnetometry measurements were performed at the Centro de Tecnologias Estratégicas do Nordeste (CETENE) under proposal number: NPSTE: PS0079/23-0032, financed by Prof. José Holanda through his research projects. The authors are grateful to the researcher John E. Pearson from the Argonne National Laboratory and Prof. Changjiang Liu from the Physics Department of the University at Buffalo - United States, for the valuable discussions on this work.

**Contributions**

L. H., J. A., F. E. analyzed all the experimental measures and J. H. discussed, wrote and supervised the work.

**Conflict of interest**

The authors declare that they have no conflict of interest.

**Referências**